# OPEN

# Optimized Spintronic Terahertz Emitters Based on Epitaxial Grown Fe/Pt Layer Structures

Garik Torosyan[1], Sascha Keller[2], Laura Scheuer[2], René Beigang[2,3] & Evangelos Th. Papaioannou[2,3]

We report on generation of pulsed broadband terahertz radiation utilizing the inverse spin hall effect in Fe/Pt bilayers on MgO and sapphire substrates. The emitter was optimized with respect to layer thickness, growth parameters, substrates and geometrical arrangement. The experimentally determined optimum layer thicknesses were in qualitative agreement with simulations of the spin current induced in the ferromagnetic layer. Our model takes into account generation of spin polarization, spin diffusion and accumulation in Fe and Pt and electrical as well as optical properties of the bilayer samples. Using the device in a counterintuitive orientation a Si lens was attached to increase the collection efficiency of the emitter. The optimized emitter provided a bandwidth of up to 8 THz which was mainly limited by the low-temperature-grown GaAs (LT-GaAS) photoconductive antenna used as detector and the pulse length of the pump laser. The THz pulse length was as short as 220 fs for a sub 100 fs pulse length of the 800 nm pump laser. Average pump powers as low as 25 mW (at a repetition rate of 75 MHz) have been used for terahertz generation. This and the general performance make the spintronic terahertz emitter compatible with established emitters based on optical rectification in nonlinear crystals.

The use of spin Hall effects to generate and manipulate spin currents has provided a large thrust in the research field of spintronics the last decade. Spin Hall effect and its reciprocal, the inverse spin Hall effect (ISHE), provide the means for conversion between spin and charge currents[1,2]. In particular, the ISHE transforms a pure spin current into a transverse charge current due to spin-orbit coupling. Extensive studies based on the ISHE have been performed since ISHE was first experimentally demonstrated by using non local detection techniques[3] and by detecting electrically pure spin currents generated in spin pumping experiments[4]. Spin pumping effect refers to the generation of a spin current from a precessing magnetization in a magnetic layer (FM). The spin current appears at the interface of the magnetic layer with a non magnetic metallic layer (NM) and it propagates in the NM layer which has to exhibit a large spin-orbit coupling like for example Pt. The ISHE in a non magnetic layer then acts as an electrical detector of the spin currents. The utilisation of ISHE in sensing spin currents has been also applied in measuring spin caloritronic effects[5,6].

Recently, the decisive role of the ISHE effect on extending the field of spintronics into the terahertz (THz) regime was revealed[7,8]. THz spintronics has the potential of application in ultra-fast current and computer technologies[9]. In the particular case of THz emission induced by the ISHE in FM/NM layers, a femtosecond laser pulse pumps a FM/NM heterostructure and generates non equilibrium spin polarized electrons in the FM layer. Subsequently, these electrons are diffused in the non-magnetic layer through a super diffusive process[10,11]. The spin current is then converted into a transient transverse charge current due to the ISHE in the NM layer. This transient current generates a short terahertz pulse that propagates perpendicular to the electrical current. The experimental demonstration of THz emission from FM/NM heterostructures due to ISHE has been beautifully shown recently[7,8,12,13]. However, the thickness dependence of the efficiency of the spintronic emitters, in particular for very small layer thicknesses has not been well understood. Seifert *et al*.[7], have shown that the THz signal exhibits a maximum for a specific total thickness of the metallic bilayer, although, according to their equation of the terahertz field amplitude[7], the signal should scale with the metal thickness. They have attributed this behavior to a resonant enhancement in a kind of a Fabry-Pérot cavity formed by the bilayer. In contrast, Yang *et al.*[13], have found no maximum in the signal by varying the Pt thickness in Fe (1.4 nm)/Pt (0–5 nm) bilayers. Instead, the signal was saturated above 3 nm Pt, possibly due to the spin diffusion length in Pt[13]. Furthermore, no pronounced maximum was observed in the Fe-thickness dependence curve for the Fe (0–5 nm)/Pt (3 nm) bilayer. Likewise, Wu *et al.*[12], reported an increase and a saturation of the peak amplitude of THz signals for a NM/Co (4 nm)/SiO$_2$ system as the thickness of the nonmagnetic layer (Pt or W) increased. By varying the thickness of the Co layer (0–10 nm) the THz peak reached a maximum value and then a gradual decrease. In this work, we present the effect of THz radiation from fully epitaxial Fe/Pt systems. We show the influence of the thickness of the individual layers on the THz emission amplitude and we discuss the role of the substrate. We demonstrate that for both Fe- and Pt-thickness dependencies the THz signal exhibits a maximum and a fast decrease after that. We quantify this behavior by using a model that takes into account generation of spin polarization starting at a minimum layer thickness of the magnetic material, spin diffusion, spin accumulation, and the electrical as well as optical properties of the bilayer samples.

[1]Photonic Center Kaiserslautern, Kaiserslautern, 67663, Germany. [2]University of Kaiserslautern, Department of Physics, Kaiserslautern, 67663, Germany. [3]University of Kaiserslautern, Research Center Optimas, Kaiserslautern, 67663, Germany. Garik Torosyan, Sascha Keller, Laura Scheuer, René Beigang and Evangelos Th. Papaioannou contributed equally to this work. Correspondence and requests for materials should be addressed to R.B. (email: beigang@physik.uni-kl.de)

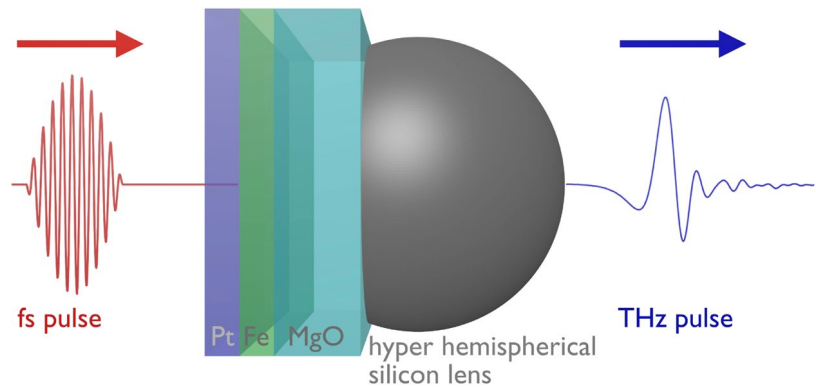

**Figure 1.** Geometrical arrangement of the samples for the generation of THz radiation. In order to collimate the strongly divergent THz beam, a hyperhemispherical silicon lens is attached directly to the substrate without damaging the delicate Pt surface.

## Results

**Geometrical arrangement of the samples.** The Fe/Pt layer structures were epitaxially grown on 0.5 mm thick MgO or Al$_2$O$_3$ substrates with Pt as the outer layer. In principle, the layer structures can be pumped from both sides, the substrate side or the side with the Pt layer. In our experiments, the layer structures were pumped from the metal side and THz emission was detected from the substrate surface. As the Pt layer thickness is in the order of 3 nm for optimized emitters most of incident pump radiation is transmitted into the magnetic layer. For experiments with thicker Pt layers the additional absorption and reflection losses have to be taken into account when comparing the experimental results with simulations. We have therefore measured directly the total absorbed pump power for the metal layer structure. Only a certain fraction of this absorbed power will be converted into spin polarized electrons in a small layer close to the Fe/Pt boundary. This has to be taken into account in our model presented below. The counterintuitive geometry has the following advantages: In order to collimate the strongly divergent THz beam a hyperhemispherical silicon lens was attached directly to the substrate with the Fe/Pt plane in the focal point of the lens without damaging the delicate Pt surface. A schematic of the structures is shown in Fig. 1.

Pumping the structure from the substrate side will always cause additional reflections from the substrate-air interface which in turn lead to strong oscillations in the corresponding THz spectra. In our geometry, the metal surface acts as an anti-reflection coating for the THz beam[14] suppressing any reflections from the substrate surface. This is illustrated in Fig. 2 for a Fe/Pt sample on a MgO substrate without a silicon lens attached. When pumped from the substrate side (upper part in Fig. 2a) a second reflected pulse from the air-substrate interface propagates in the direction of the main pulse. Due to the high index of refraction of MgO at THz frequencies (n = 3) about 50% of the amplitude of the backward pulse is reflected. It can clearly be seen that there are no additional reflections causing oscillations in the THz spectra when pumping the structure from the metal side (see the lower part of Fig. 2b). All further experiments were done in this counterintuitive geometry with the silicon lens attached if not stated otherwise. Because of nearly index matching between MgO and silicon there is almost no loss at the MgO silicon interface. With this set-up the samples were easily and reproducibly changed without destroying the alignment of the system allowing for a quantitative comparison between different samples (see Methods section).

**Pulse length and bandwith.** The pulse length and bandwidth of the THz pulse strongly depend on the rise and fall time of the transient electrical current in the nonmagnetic layer induced via the ISHE. Whereas the rise time is mainly determined by the pump pulse length and the diffusion properties of the spin current, the fall time is limited by the relaxation time of the electrons in the conducting material. For the relaxation time we have considered a superdiffusive transport process according to studies by M. Battiato[15] resulting in a relaxation time in the order of 5 fs in Pt[16]. Such a relaxation time supports bandwidths up to several tens of THz and does not limit the bandwidth we have observed in our experiments.

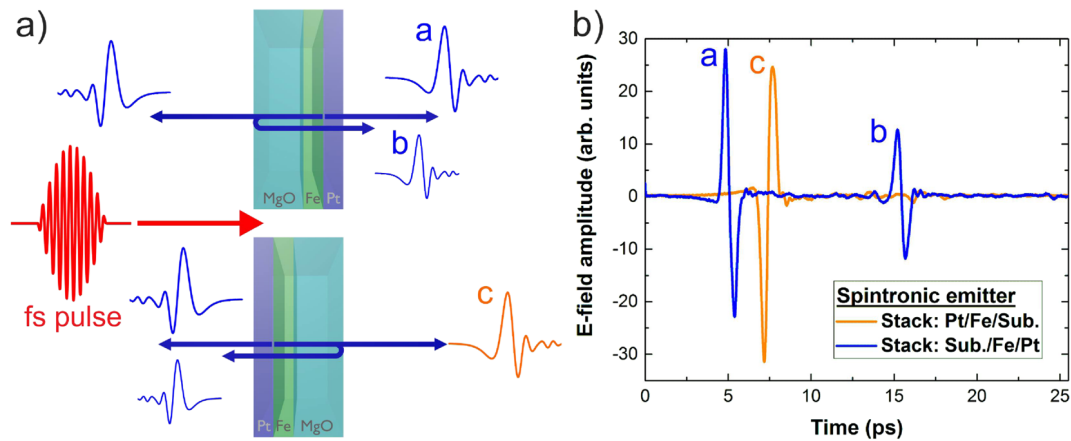

**Figure 2.** (**a**) Orientation of the sample with respect to the pump beam and (**b**) the generated THz pulses. In the orientation with the metal side towards the pump beam the reflection from the MgO-air interface is suppressed.

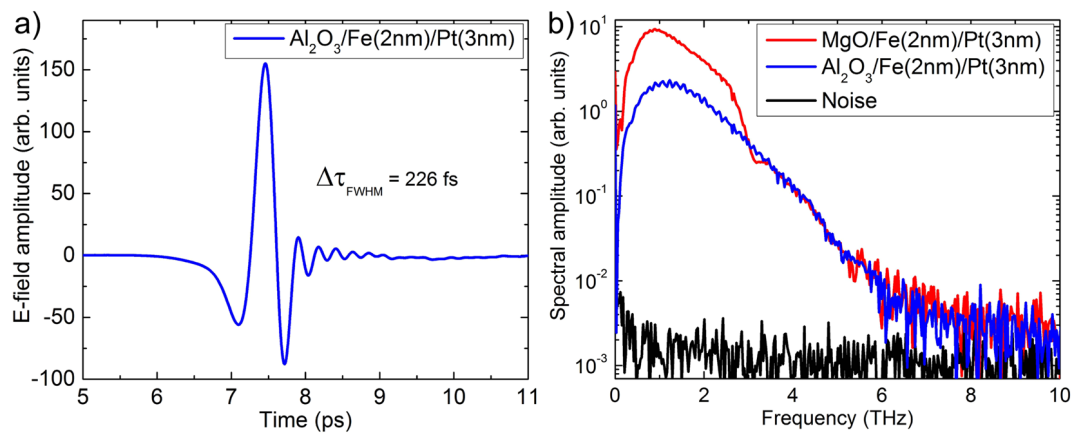

**Figure 3.** (**a**) Typical THz pulse obtained from a Fe(2 nm)/Pt(3 nm) sample. (**b**) Spectra of the generated THz pulses for samples with two different substrates, MgO (red) and $Al_2O_3$ (blue).

In addition, absorption in the layer structures and the substrate influence both pulse length and bandwidth. A typical pulse and the corresponding spectra are shown in Fig. 3 for a 2 nm/3 nm Fe/Pt layer structure on MgO and $Al_2O_3$, respectively. The minimum pulse width was measured to be below 250 fs for both substrates with a tendency to shorter pulses for the $Al_2O_3$ substrate (see Fig. 3a). Spectra for bilayers on both substrates are shown in Fig. 3b) extending to approximately 8 THz. All spectra are shown as measured and are not corrected for the detector response. Above 3 THz the well known strong THz absorption of MgO is visible for the MgO substrate. The dynamic range is well above 60 dB with a maximum around 1.5 THz. This is comparable with spectra obtained from photoconductive emitters. The maximum frequency measured in these experiments was finally determined by the frequency response of the dipole antenna of our photoconductive switch which was used as the detector (see Methods section). The pulse length of 50 fs of our pump source would allow for even broader spectra. Furthermore, the absorption around 8 THz in GaAs which is used as the semiconductor material for the detector antenna limits the bandwidth. Having the choice between MgO and $Al_2O_3$ substrates MgO provides the higher dynamic range at frequencies below 3 THz. This is probably caused by the fact that there is an almost perfect epitaxial growth of Fe on MgO whereas on $Al_2O_3$ it is not the case, resulting in a smaller dynamic range. This finding is supported by results of experiments with polycrystalline Pt layers on Fe/MgO resulting in considerably weaker THz signals. Together with a different detector (e. g. using electro-optical sampling in GaP) and even shorter pump pulses a much broader bandwidth can be obtained. This has been shown recently by Kampfrath et al.[7].

**Thickness dependence.** In order to optimize our THz emitter we have performed a systematic study of the dependence of the THz amplitude on the thickness of the Pt and Fe layers for samples epitaxially grown on 0.5 mm thick MgO substrates. In a first set of experiments the Fe layer thickness was kept constant to 12 nm whereas the Pt layer thickness was varied from 0.25 nm to 12 nm. In the second set the Pt layer thickness was fixed at 3 nm while the Fe layer thickness was changed from 1 nm to 12 nm. The results are shown in Fig. 4. In Fig. 4a) the Fe (12 nm)/Pt (x nm) series is shown. As a measure of the strength of the THz signal, the peak to peak amplitude of the first two oscillations of the THz pulse was determined.

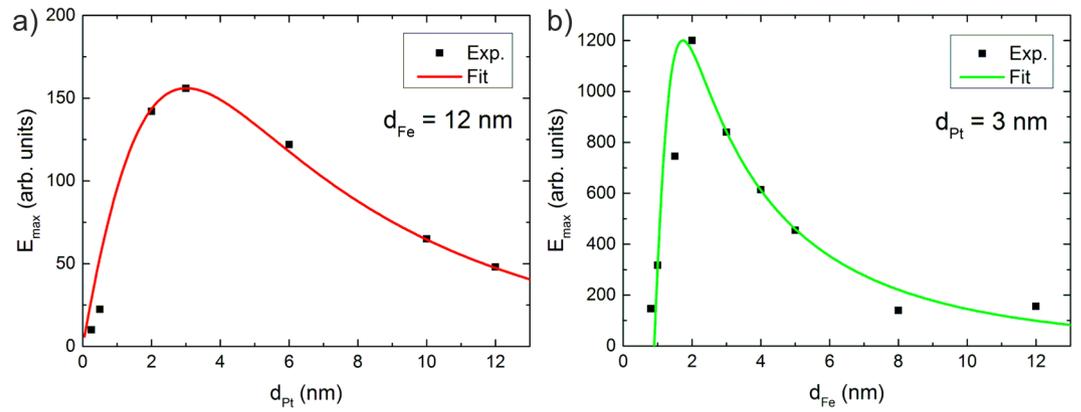

**Figure 4.** (**a**) Pt thickness dependence of the THz field amplitude for a constant Fe thickness of 12 nm. (**b**) Fe thickness dependence of the THz field amplitude for a constant Pt thickness of 3 nm that has provided the highest signal in (**a**).

The variation in pump beam absorption through Pt layers with changing thicknesses has been taken into account for the comparison. The theoretical fitting curve in Fig. 4a) and b) is based on the following equation 1:

$$E_{THz} \propto \frac{P_{abs}}{d_{Fe} + d_{Pt}} \cdot \tanh\left(\frac{d_{Fe} - d_0}{2\lambda_{pol}}\right) \cdot \frac{1}{n_{air} + n_{MgO} + Z_0 \cdot (\sigma_{Fe} d_{Fe} + \sigma_{Pt} d_{Pt})} \cdot \tanh\left(\frac{d_{Pt}}{2\lambda_{Pt}}\right)$$
$$\cdot \, e^{-(d_{Fe} + d_{Pt})/s_{THz}} \tag{1}$$

where $n_{air}$, $n_{MgO}$ and $Z_0$ are the index of refraction of air, the index of refraction of MgO at THz frequencies and the impedance of vacuum, respectively.

Equation 1 corresponds to a model that describes the dependence of the generated THz signal on layer thicknesses of Fe and Pt layers. In particular, it describes the dependence on the individual layer thicknesses and not only on the total layer thickness of Fe and Pt layers. The model explains the onset of THz generation above a certain Fe layer thickness and the maximum at relatively small layer thicknesses. It takes into account all successive effects after the laser pulse impinges on the bilayer, including the absorption of the fs pulse in the Pt and Fe layers, the generation and diffusion of spin currents in Fe and Pt as well as the generation of THz radiation and its attenuation in the metal layers. In detail, the first term accounts for the absorption of the femtosecond laser pulse in the metal layers. As only spin polarized electrons within a certain distance from the boundary between Fe and Pt will reach the Pt layer only a fraction of the measured total absorbed power contributes to the generated THz signal. This fraction scales with the inverse of the total metal layer thickness $1/(d_{Fe} + d_{Pt})$. The second term describes generation and diffusion of the generated spin current flowing in Fe towards the interface with Pt; to elaborate the possibility that extra thin Fe layers can loose their ferromagnetic properties below a certain thickness we introduce the term $d_0$ in equation 1. Below this critical thickness we consider that the flow (if any) of spin current in Fe does not reach the Pt layer. Above this critical thickness the generated spin polarization saturates with a characteristic constant $\lambda_{pol}$. Magneto-optical Kerr effect measurements confirm this assumption (see Methods section).

The third and forth term, the tangent hyberbolic function divided by the total impedance, refers to the spin accumulation in Pt which is responsible for the strength of the THz radiation and it depends on the finite diffusion length $\lambda_{Pt}$ of the spin current in Pt. The symbols $\sigma_{Fe}$ and $\sigma_{Pt}$ are the electrical conductivities of the two materials, respectively. This part of the equation is according to the theory of spin pumping effect[17,18] and includes the shunting effect of the parallel connection of the resistances of the individual Fe and Pt layers. Although the electrical conductivity depends on layer thickness for very thin layers we have used a constant value for the layer thicknesses. This value is considerably smaller than the bulk value as in the thickness ranges in our experiments the bulk value has not been reached[19].

The last term describes the attenuation of the THz radiation during propagation through the metal layers (with $s_{THz}$ as an effective inverse attenuation coefficient of THz radiation in the two metal layers). In the case of small losses the attenuation can be taken into account by this single exponential factor and the third factor which accounts for the multiple reflections of the THz pulse at the metal/dielectric interfaces (see Methods section). All terms together describe the layer thickness dependence of the measured THz amplitudes. The influence of the silicon lens on the thickness dependence of the generation process can be neglected as the distance between lens and metallic layers is five orders of magnitude larger than the layer thicknesses itself. Experiments with and without silicon lens resulted in the same thickness dependence.

In Fig. 4a there is rather good qualitative agreement between our experimental data and the theoretical expectation. For the simulations we have assumed a spin current diffusion length in Pt of $\lambda_{Pt} = 1.40$ nm. This is in good agreement with direct measurements of the diffusion length using microwave techniques[20]. The other constants used for this fit are $n_{air} = 1$, $n_{MgO} = 3$, $Z_0 = 377\,\Omega$, $\sigma_{Fe} = 6 \cdot 10^6\,\Omega^{-1}m^{-1}$, $\sigma_{Pt} = 2 \cdot 10^6\,\Omega^{-1}m^{-1}$, $d_0 = 0.9$ nm, $\lambda_{pol} = 0.3$ nm, and $s_{THz} = 22$ nm. The optimum Pt layer thickness for maximum THz generation seems to be between 2 and 3 nm.

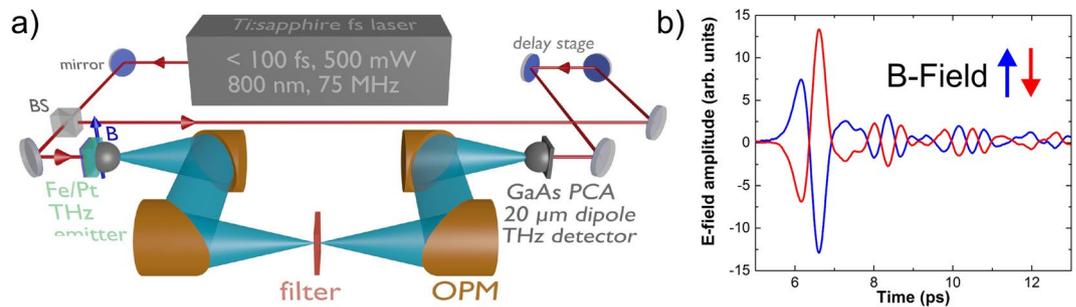

**Figure 5.** (**a**) Experimental set-up for THz generation. (**b**) THz pulses for two opposite directions of the magnetic field.

We have used this optimum Pt layer thickness for the determination of the optimum Fe layer thickness. With a fixed Pt layer thickness of 3 nm the Fe layer thickness was changed from 1 nm to 12 nm. The results are shown in Fig. 4b again together with the simulation using the same parameters as for the previous measurements. For the comparison the variation in pump light absorption in the magnetic layer has been accounted for. There is a steady increase in THz power with decreasing Fe layer thickness following the theoretical simulations. However, for a Fe layer thickness below 2 nm there is a considerable decrease in THz amplitude which cannot be explained alone by the decreasing pump light absorption. We attribute this decrease in signal to the onset of loss of magnetic properties of the Fe layer resulting in a critical thickness $d_0 = 0.9$ nm. As a consequence, the optimum layer structure for maximum THz generation is 3 nm of Pt on 2 nm of Fe.

## Discussion

We have shown that Fe/Pt epitaxial bilayer systems can be considered as a competitive THz radiation source of spintronic nature. The spintronic THz emitter based on the ISHE presented here has a number of advantages compared to other optical and electro-optical THz emitters: The alignment of the emitter with respect to the pump beam is simplified as one has a free choice choosing the focusing spot within the sample surface. A collimating Si lens can be attached directly to the substrate of the bilayer sample collecting most of the strongly divergent THz beam. Thus, changing the NM/FM bilayer samples becomes possible without loss of alignment of the total THz system. Up to pump fluences of 5 mJ/cm$^2$ we did not observe any damage of our samples (see Methods section). There are no electrical connections required for the operation of the emitter. The polarization of the generated THz radiation can easily be changed by changing the weak magnetic field perpendicular to the direction of the pump beam. The average pump power levels required to generate useful THz signals is comparable to power levels used in optical rectification methods in nonlinear crystals. As the excitation of electrons in the Fe layer is independent of wavelength other pulsed laser sources can be used which are easier to operate like e.g. fs fiber lasers at 1.5 $\mu$m. The THz pulse length and bandwidth is, in principle, only limited by the relaxation time of hot electrons in the Pt material. The best configuration of the THz emitter was revealed by varying Fe and Pt layer thicknesses. Samples with 2 nm Fe and 3 nm Pt gave the maximum THz amplitude. To quantify the thickness dependence of the THz amplitude we have successfully developed a model that takes into account the optical absorption in the metal layers, the generation and diffusion of hot carries in Fe, the shunting effect, spin accumulation in Pt and the THz absorption in Fe. The optimization and the modeling of the radiation with the materials properties is necessary for future applications of the effect.

## Methods

**THz time domain set-up.** A standard terahertz time-domain spectroscopy (THz-TDS) system, described in detail elsewhere[21], has been used for generation and measurements of THz waveforms from different spintronic emitters (see Fig. 5). The system is driven by a femtosecond Ti:Sa laser delivering sub-100 fs optical pulses at a repetition rate of 75 MHz with an average output power of typically 600 mW. The laser beam is split into a pump and probe beam by a 90:10 beam-splitter. The stronger part is led through a mechanical computer-controlled delay line to pump the THz emitter, and the weaker part is used to gate the detector photoconductive switch with a 20 $\mu$m dipole antenna. In a classical (standard) THz-setup both the emitter and the detector operate with photoconductive antennas (PCA) whereas in the present work the PCA emitter is substituted by a spintronic (ST) Fe/Pt bilayer sample which is placed in a weak magnetic (20 mT) field perpendicular to the direction of the pump beam and in the direction of the easy axis of Fe. The direction of the magnetic field determines the polarization of the THz field which is perpendicular to the direction of the magnetic field. Changing the direction of magnetic field into the opposite direction changes the phase of the detected THz waveforms by 180° (Fig. 5b). In this way, by changing the orientation of the magnetic field the polarization of the generated THz radiation can be changed easily.

The optical pump beam that is sharply focused onto the sample at normal incidence by an aspherical short-focus lens, excites spin polarized electrons in the magnetic layer (Fe) which give rise to a spin-current, which in turn excites a transverse transient electric current in the Pt layer. The latter results in THz pulse generation of sub-picosecond duration being emitted forward and backward into free space in the form of a strongly divergent beam. The central wavelength within the beam is about 200 μm and is much longer than the diameter of its source which is smaller than 10 μm. That is why the emitted THz field would fill the half-space behind the sample. A hyperhemispherical Si-lens being attached to the back side of the sample collects it into a divergent beam in the form of a cone and directs it further. With the lens attached an enhancement factor of up to 30 in electric field amplitude has been observed. After the lens the beam is collimated with an off-axis parabolic mirror and sent to an identical parabolic mirror in the reversed configuration. The latter focuses the beam to the second Si-lens which finally focuses the beam through the GaAs substrate of the detector PCA onto the dipole gap for detection. In this way the THz optical system consisting of the two Si-lenses and the two parabolic mirrors images the point source of THz wave on the emitter surface onto the gap of the detector PCA and ensures an efficient transfer of the emitted THz emission from its source to the detector. In the present paper a 4-mirror THz-optics is used making it possible -besides two collimated beam paths- also to have an intermediate focus of the THz beam aimed at imaging applications (Fig. 5a)). In our set-up the THz beam path is determined by the silicon lens on the emitter, the parabolic mirrors and the silicon lens on the photoconducting antenna of the detector. The alignment of these components is not changed during an exchange of the spintronic emitter. If in addition the position of the pump beam focus remains constant the spintronic emitter can easily be exchanged without changing the beam path as the lateral position of the focus on the emitter is not critical assuming a homogeneous lateral layer structure. The frequency response of the photoconductive dipole antenna with 20 μm dipole length limits the observable bandwidth. With this dipole length a maximum in the detector response around 1 THz can be expected with a reduction to 50% at 330 GHz and 2 THz. The 10% values are at 100 GHz and 3 THz. This estimation is based on calculations by Jepsen *et al.*[22]. Above 3 THz the frequency response of the detector is very flat and at 8 THz a strong phonon resonance in GaAs which is used as substrate material for the photoconductive antenna causes strong absorption of the THz radiation. Above 8 THz no THz radiation was detected.

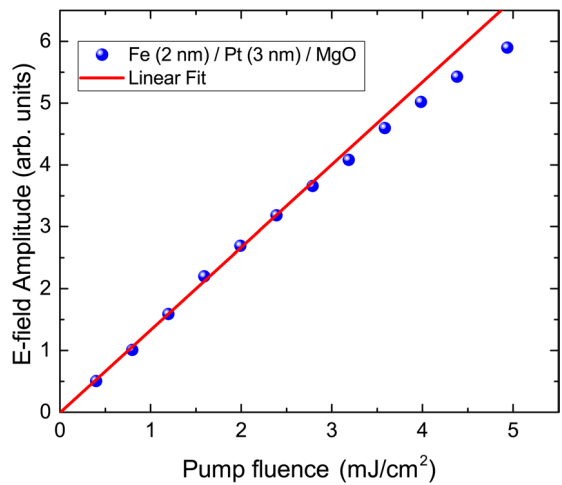

**Figure 6.** Maximum E-field amplitude as a function of pump fluence.

The delay line provides for the synchronous arrival of the weaker part of the optical pulse and that of the THz pulse at the detector antenna gap from either side, as well as for scanning in time the "open" state of the gap along the THz pulse duration. In each position of the delay line the transient current induced in the detector by the electric field of the THz sub-picosecond pulse is proportional to its instantaneous electrical field value. It is summed up from many laser pulses which reach the detector during the single step of the delay line and integrated within the "open" state time window. It is measured as one single point of the THz wave-form and is in the order of several nano amperes at the maximum of the THz pulse. Hence, for its measurement lock-in technique has to be used. For that purpose the pump beam is mechanically chopped at 1.35 kHz frequency. By scanning the "open" state of the detector in time the THz pulse shape can be sampled. Utilizing the magnetic field dependence of the THz polarization the THz beam can, in principle, be chopped electrically by means of an alternating magnetic field around the emitter.

In order to compare our results with results from other THz emitters we have characterized the generated THz radiation using the maximum dynamic range which is the maximum spectral amplitude above noise level and the maximum frequency above noise level. The last quantity, of course, grows with measurement time up to a maximum measurement time limited by the stability of the whole system. For our measurements we have used a scan range of 33 ps with a step width of 0.4 μm of our delay line, a scan speed of 30 μm/s and an integration time of 100 ms.

We have measured the THz amplitude as a function of pump power for a fixed spot size of the pump beam. With a spot size of 5 μm radius, a repetition rate of 75 MHz of our pump laser and a maximum average output power of 350 mW we changed the pump fluence on the sample from 0.5 mJ/cm$^2$ to approximately 5 mJ/cm$^2$ without any damage of the sample. The slight onset of saturation of the THz signal may be caused by a temperature increase of the sample (see Fig. 6).

**Sample preparation-Epitaxial growth.** Fe thin films were grown epitaxially on the 0.5 mm-thick MgO (100) and $Al_2O_3$ (0001) substrates by molecular beam epitaxy (MBE) technique in an ultrahigh vacuum (UHV) chamber with a base pressure of $3 \times 10^{-10}$ mbar. The growth rate was R = 0.05 Å/s controlled by a quartz crystal during the deposition procedure. The incident Fe beam was perpendicular to the substrates. The cleaning protocol of the MgO (001) $1 \times 1 cm^2$ substrates involved heating at 600 °C for 1 hour. The deposition of Fe was performed at 300 °C temperature. At a next growth stage, on top of the Fe layer, a Pt layer was deposited at the same temperature. The thickness of the individual layers has been varied for Fe (1–12 nm) and for Pt (1–18 nm) as it was monitored *in-situ* by a calibrated quartz crystal oscillator and confirmed *ex-situ* by X-ray reflectivity (XRR) measurements. X-rays diffraction (XRD) measurements have shown for the thicker samples (Fe and Pt thicknesses greater than 2 nm) the presence of Fe (002) peaks revealing the epitaxy of Fe layer on the MgO substrate, with Fe(001)//MgO (011) for all samples. Diffraction peaks of Pt (200) and Pt (400) arising from the Pt top layer parallel to the Fe (200) planes were observed revealing the epitaxial relation of Pt on Fe. The growth of the fcc Pt layer on bcc Fe along the [100] plane direction is correlated to the Bain epitaxial orientation[23,24] when the Pt cell is 45 degrees rotated with respect to the Fe lattice. The structural quality of the Fe/Pt bilayers was further investigated by measuring in-plane XRD polar plots and selectively transmission electron microscopy (TEM). Both methods confirmed the epitaxial quality of the bilayers. The growth of Fe on sapphire (0001) is not perfect epitaxial because of a larger mismatch in lattice constants. This may also influence the THz generation. In our experiments we have observed a weaker spectral amplitude at the maximum frequency whereas there is no obvious absorption at higher frequencies.

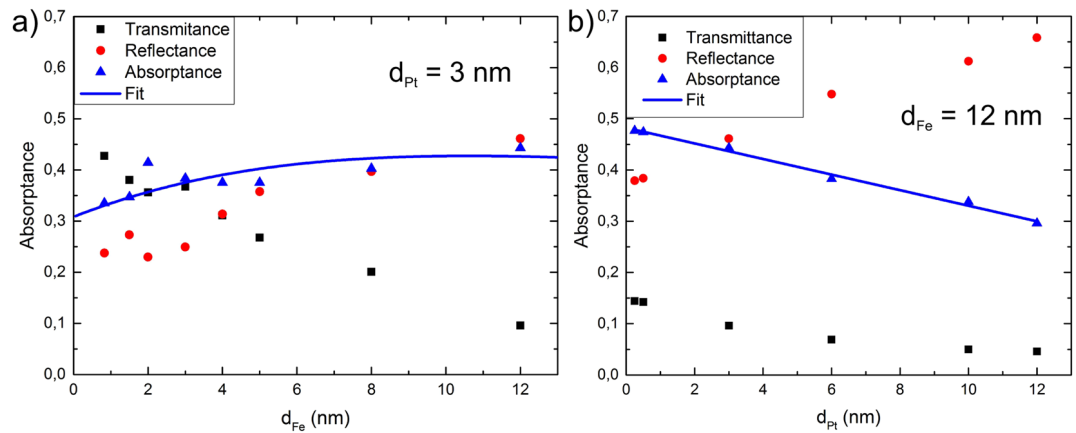

**Figure 7.** Transmittance, reflectance (measured) and absorptance (calculated) of the Fe/Pt emitters grown on MgO substrates.

**Theoretical model.** The charge current $j_c$ in Pt responsible for THz generation depends on the generated spin polarization in Fe, the diffusion of spin polarized electrons in Fe and the diffusion of spin polarized electrons in Pt which are converted into a charge current via the inverse spin Hall effect.

The first term in equation 1 takes into account the absorption in the metal layers when the sample is pumped from the metal layer side which reduces the power available for generation of spin polarized electrons in Fe. As only spin polarized electrons within a certain distance from the boundary between Fe and Pt will reach the Pt layer, only a fraction of the measured total absorbed power contributes to the generated THz signal. This fraction scales with the inverse of the total metal layer thickness $1/(d_{Fe} + d_{Pt})$ assuming a constant diameter of the pump beam and an almost linear dependence of the absorbed power on layer thickness.

$$j_c \propto \frac{P_{abs}}{d_{Fe} + d_{Pt}} \quad (2)$$

Both assumptions are valid in our case for layer thicknesses investigated in our experiments. Typical absorption measurements are shown in Fig. 7 using

$$T = P_{trans}/P_{in}, \quad R = P_{refl}/P_{in}, \quad A = P_{abs}/P_{in} \quad (3)$$

Where $P_{in}$ is the incoming laser power focused on to the emitter, $P_{trans}$ is the laser power transmitted by the emitter and $P_{abs}$ the laser power absorbed by the emitter. To determine the absorptance A we have measured R and T and use $A = 1 - T - R$. It can be seen that for small metal thicknesses the absorptance dominates whereas for larger metal layer thicknesses the reflection is dominant.

The very steep onset of THz radiation at very thin Fe layer thicknesses above a certain minimum Fe thickness is described by the second term in equation 1. It describes the development of spin polarization starting at a layer thickness above a certain "dead layer" $d_0$ and increases with a certain constant $\lambda_{pol}$ which is characteristic for the saturation of spin polarization with layer thickness (see e.g. Allenspach *et al.*)[25].

In order to support this assumption we have studied the magnetic properties of our sample series with the help of longitudinal magneto-optic magnetometry. The results are illustrated in Fig. 8. The variation of hysteresis loop characteristics as a function of Fe layer thickness demonstrates that the samples with thickness larger than 1.5 nm have the easy axis of the magnetization in-plane. The sample with 0.8 nm (2.8 monolayers (ML)) has a very small signal and exhibits a loop that resembles a superparamagnetic behaviour of Fe indicative that the sample could be close to its Curie point. The latter is in agreement with other references about the Curie temperature of 2–3 ML of Fe layer thickness on different substrates[26]. In such case the "dead" magnetic layer can be estimated to be below 0.8 nm. This is also in agreement with results obtained by Allensbach et al.[25] where for thicknesses up to 2 ML of Fe on Cu (001) no spin polarization was detected. For the case of a few ML of Fe there is also the possibility that the spin polarization has a perpendicular component[26]. Unfortunately, we have no experimental access to measure the out-of-plane component of the magnetization. However, the presence or non-presence of perpendicular magnetization is not going to change our results since our THz signal is sensitive to the in-plane component of the magnetization.

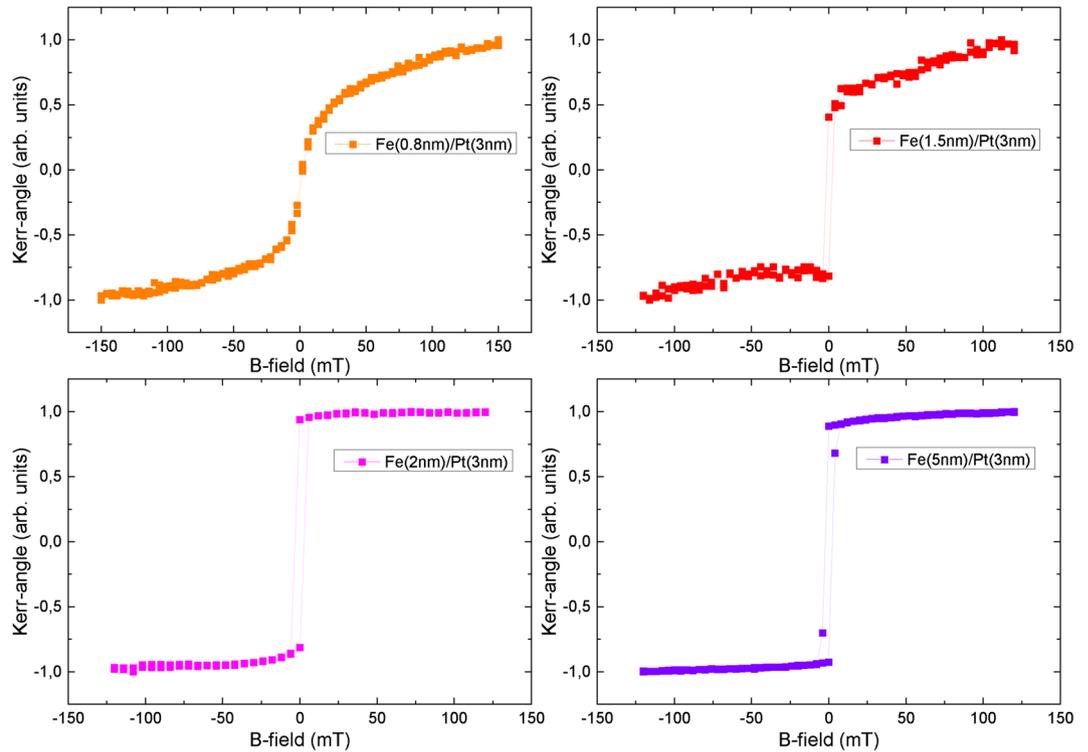

**Figure 8.** Magneto-optical Kerr effect measurements of the Fe/Pt emitters with Fe thicknesses ranging from 0.8 to 5.0 nm and constant Pt thickness of 3 nm.

Regardless of the reason (either close to $T_c$ or no in-plane component due to the perpendicular orientation of the magnetization) the MOKE data justify our assumption of a "dead" magnetic layer for the THz emission below 0.8 nm of Fe.

The third and forth term, the tangent hyberbolic function divided by the total impedance, refers to the spin accumulation in Pt which is responsible for the strength of the THz radiation and it depends on the finite diffusion length ($\lambda_{Pt}$) of the spin current in Pt. This part of the equation is according to the theory of generation of THz radiation from a current source based on the inverse spin Hall effect[8,17,18].

The last term accounts for losses in the metal layers. It can be described by a single exponential attenuation factor under the following assumptions: Deriving the generated THz signal via the ISHE using Green's function only negligible losses were assumed and a real wave vector k was used for the THz propagation in the metal layers (see e. g. Seifert et al.)[8]. Taking into account small losses for the generated THz radiation and replacing the real wave vector $k = \frac{n \cdot \omega}{c}$ by a complex wave vector $k^* = \frac{n^* \cdot \omega}{c} = \frac{n \cdot \omega}{c} - i \cdot \frac{\kappa \cdot \omega}{c}$ results in an additional attenuation factor $e^{-\frac{\kappa \cdot \omega}{c} \cdot z}$ where z is the propagation distance in metal, $\kappa$ the imaginary part of the index of refraction, $\omega$ the angular frequency, and c the vacuum speed of light, respectively. Assuming only small losses the attenuation factor is independent from z and the E field is constant across the metal layer if the thickness is small compared to typical attenuation lengths. Following the derivation of Seifert et al[8]. this factor becomes a constant factor which is independent from other parameters and only depends on the total thickness of the metal layers. The smaller the layer thicknesses the better is this approximation. Due to multiple reflections in the metal (accounted for by the third term in equation 1) the effective propagation length in metal is larger than the sum of the layer thicknesses of Pt and Fe and depends on the finesse of the metal Fabry-Perot structure. As this value is not known exactly an effective inverse absorption constant $s_{THz}$ was used for the fit and the propagation length was set to the sum of Pt and Fe layer thicknesses. This value is smaller than the real inverse absorption constant in metal as experimentally determined by Yasuda and Hosako[27] (typically $s_{THz}$ = 150 nm for gold averaged over the spectral range from 0.5 THz to 3 THz). To get the real attenuation length our fitted effective inverse absorption constant has to be multiplied by the finesse.

The influence of the substrate is neglected in our model as the attenuation of the THz radiation in the substrate is constant for different layer thicknesses and we are discussing only the dependence on Fe/Pt layer thicknesses. The influence of the substrate can be seen in the spectral amplitude. Above 3 THz there is a strong THz absorption which reduces the THz bandwidth. There is of course a different THz absorption in different substrates which has to be taken into account when comparing different substrates.

## Acknowledgements

We thank Burkard Hillebrands for his scientific support. E.Th.P. and S.K. acknowledge support from the Deutsche Forschungsgemeinschaft (DFG) through the collaborative research center SFB TRR 173: SPIN+X Project B07 and the Carl Zeiss Foundation. We thank Jörg Lösch from the Institute of surface-analytics (IFOS-Kaiserslautern) for his support with the XRR-XRD measurements. We thank Marco Battiato and Hans Christian Schneider for helpful discussions.

## Author Contributions

G.T. and L.S. conducted the experiments, S.K. and L.S. grew the samples, R.B. and E.Th.P. conceived the experiments, R.B., G.T., S.K. and E.Th.P. analyzed the results. All authors reviewed the manuscript.

## Additional Information

**Competing Interests:** The authors declare that they have no competing interests.

**Publisher's note:** Springer Nature remains neutral with regard to jurisdictional claims in published maps and institutional affiliations.